\documentclass[prl,aps,twocolumn,superscriptaddress]{revtex4}

\usepackage{epsfig}
\usepackage{amsmath}

\newcommand{\be}{\begin{equation}}
\newcommand{\ee}{\end{equation}}
\newcommand{\bea}{\begin{eqnarray}}
\newcommand{\eea}{\end{eqnarray}}
\newcommand{\la}{\langle}
\newcommand{\ra}{\rangle}

\newcommand{\HH}{{\cal H}}

\newcommand{\Tr}{{\mathrm{Tr}}}
\newcommand{\No}{{\hat N}_{\mathrm{odd}}}
\newcommand{\Nodd}{{N}_{\mathrm{odd}}}
\newcommand{\ket}[1]{{|{#1}\ra}}
\newcommand{\bra}[1]{{\la{#1}|}}
\newcommand{\e}{{\mathrm e}}
\newcommand{\im}{{\mathrm i}}
\newcommand{\nmax}{n_{\mathrm{max}}}

\usepackage{ulem}
\usepackage{color}
\definecolor{darkblue}{rgb}{0.0,0,0.6}
\definecolor{darkred}{rgb}{0.6,0,0.0}

\begin{document}

\author{Peter Schmitteckert}
\affiliation{DFG Center for Functional Nanostructures, Karlsruhe Institute of Technology, 76128 Karlsruhe, Germany}
\affiliation{Institute of Nanotechnology, Karlsruhe Institute of Technology, 76344 Eggenstein-Leopoldshafen, Germany}

\title{On the relaxation towards equilibrium in an isolated strongly correlated 1D Bose gas}

\date{\today}


\begin{abstract}
In this work we study the time evolution of soft core bosons on a one-dimensional lattice, where the particles
are initially quenched into a atomic density wave. At time $t=0$ the particles are released from the quench and
can evolve under the dynamics of a soft-core Bose-Hubbard Hamiltonian on a lattice including a confining trap. 
\end{abstract}

\maketitle

Ultra cold atoms in optical lattices have become a testbed to compare simulations for strongly correlated quantum systems
with experimental realizations\cite{Jaksch_Zoller:AP2005,Sen:AdvPhys2007,Bloch:RMP2008}.
In a recent work Trotzky et al.\ \cite{Trotzky:} reported on the experimental realization and the numerical simulation 
of the dynamics of a quench of a strongly interacting Bose gas in one dimension. 
From their work they draw the remarkable conclusion that
``for intermediate times the system fulfills the promise of being a dynamical quantum simulator, in that the controlled dynamics runs for
  longer times than present classical algorithms based on matrix product states can efficiently keep track of''.
In this work I present numerical simulations based on time dependent DMRG as described in \cite{PS:PRB04}
for sample 2c of ref.\cite{Trotzky:} to demonstrate that simulations on classical computers based on matrix product states
can performed reliably on time scale which exceeds the time scale of the reported experimental data.%
\footnote{The actual td-DMRG variant used in \cite{Trotzky:} is not stated in the paper.
Presumably they used an adaptive td-DMRG method combined with a Trotter-Suzuki based decomposition of the time evolution operator\cite{White-2004}}.

In the experiment under consideration a one dimensional Bose gas in a harmonic trap was subject to two optical lattices.
The first optical lattice was used to provide a lattice, which can be modeled by the Hamiltonian
\begin{equation}
 \HH = \sum_x \left( \hat{a}^+_{x-1} \hat{a}^+_x + \text{h.c.} \right) + \frac{U}{2} \hat{n}_{x-1} (\hat{n_x} - 1) + \sum_x \frac{K x^2}{2} \hat{n}_x \label{eq:H}
\end{equation}
where $\hat{a}^{}_x$ ($\hat{a}^{+}_x$) annihilates (creates) a particle at site $x$, $\hat{n}_x = \hat{a}^{+}_x \hat{a}^{}_x$ gives the number
of particles at site $x$, $U$ is the on-site interaction and $K$ is the potential of harmonic trap. In the following we use the parameter of sample (c)
of reference~\cite{Trotzky:}: $U/J=5.16$, $K/J=9\cdot10^{-3}$, the number of lattice sites $M=121$ with $N = \sum_x \la \hat{n}_x  \ra= 43$ particles,
and the system is centered symmetrically around $x=0$.
We have set $J=1$ for convenience. 
The second optical lattice is used to quench the particles on odd sites only for time $t<0$ and is switched off at $t=0$ so that
the particles can now propagate between the odd an even sites and thefore through the complete system.
In order to model the physical situation we start with an eigenstate of an Hamiltonian consisting of a staggered local potential,
a strong on-site repulsion and a small coupling $J$, see below, whicj is calculated via 
a standard density matrix renormalization group (DMRG)\cite{DMRG}.
In order to perform the time dependent simulations we applied the full td-DMRG\cite{PS:PRB04}.
Specifically, our simulation consist of the following steps:

\begin{itemize}
 \item First we perform a ground state infinite lattice sweep as a warm up.
 \item We perform 9 finite lattice sweeps where we target for the ground state $\ket{\Psi_0}$ of an Hamiltonian $\HH_0 $, $\HH_0 \ket{\Psi_0} = E_0 \ket{\Psi_0}$.
 \item At each DMRG step we perform a time evolution of $N_t$ time steps of size $\Delta_t$: $\ket{\Psi_{t_{n+1}}} = \e^{-\im (\HH -E) \Delta_t} \ket{\Psi_{t_{n}}}$,
       $t_n = n \Delta_t$, $E= \bra{\Psi_0} \HH \ket{\Psi_0}$.
       The density matrix used to select the basis states kept is given by the mixed density matrix $\rho = \Tr \sum_{n=0}^{N_t} \ket{\Psi_{t_{n}}}\bra{\Psi_{t_{n}}}$,
       that is, in each DMRG step we include the complete time evolution to select the target space.
 \item The action of the matrix exponential $\e^{-\im (\HH -E) \Delta_t} $,
       is evaluated via a Krylov sub space expansion. There one expands
       the matrix exponential in the Krylov space ${\cal K}^{m}_\HH (\ket{\Psi}) = {\mathrm{span}}\left\{ \ket{\Psi}, \HH \ket{\Psi}, \cdots, \HH^{m-1}\ket{\Psi}\right\}$.
       The accuracy of this expansion is comparable to exact diagonalization using sparse matrix methods.\footnote{We actually use the full Arnoldi procedure\cite{Saad:SIAM1992,Moler_VanLoan:SIAM2003}.
        While this is formally equivalent to a Lanczos version, where one only uses the tridiagonal elements of $\HH$ projected
        on the Krylov space. We found that on finite precision arithmetic the full Arnoldi provides a higher accuracy.}
        Here we use a minimal residual of $10^{-10}$ for
        the accuracy of the matrix exponential.\cite{Saad:SIAM1992,Moler_VanLoan:SIAM2003} 
 \item  Once the initial short time dynamics is finished we continue the above described finite lattice DMRG sweeping, where we then increase the 
        number of time steps and the number of states kept per block. By restarting the DMRG at a given number of time steps and increasing
        the number of states $m$ kept per block, we can actually check for convergence. In our DMRG algorithm we always used   
        a $A \bullet\bullet B$ blocking scheme, where $m$ counts only the number of states kept per blocks $A$, $B$. The inserted sites are not
        included in $m$.
 \item  In order to perform simulations for soft core bosons we have to restrict the maximal occupation $\nmax$ of a given site, where
       we use up to $\nmax=5$.
\end{itemize}

\begin{figure}
\begin{center}
\includegraphics[width=3in,clip=true]{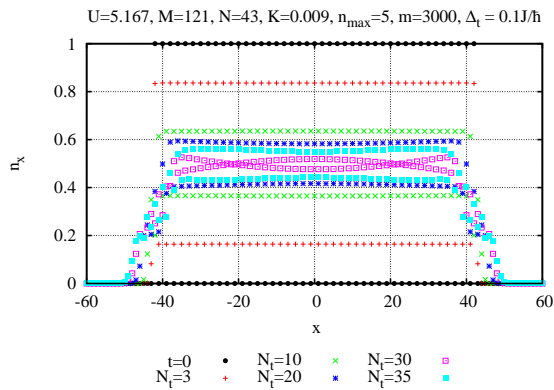}
\end{center}
\caption{Snapshots of the evolution of the initial atom density waves at $t=0$ (circles) under a Bose Hubbard Hamiltonian $\HH$ 
 on a $M=121$ site system filled with $N=43$ particles applying an on site repulsion $U=5.167$, a trap potential $K=0.009$, and a hopping element
  of $J=1$. In the td-DMRG we allowed for an maximal local site occupation of $\nmax=5$, used 3000 states per DMRG block and applied a time step
  of $\Delta_t= 0.1J/\hbar$.} \label{fig:nx}.
\end{figure}

In Fig.~\ref{fig:nx} we show the evolution of the initially quenched atomic density wave on $M=121$ lattice sites consisting of $N=43$ particles
in the center of the trap, where only the odd sites are occupied with a single particle at time $t=0$. 
Within the time scale of the simulation, the system stays roughly homogeneous in the center of the trap --- except the odd/even oscillation ---
and displays an expansion at the border of the particle cloud, which does not yet reach the boundary of the system.

\begin{figure}
\begin{center}
\includegraphics[width=3in,clip=true]{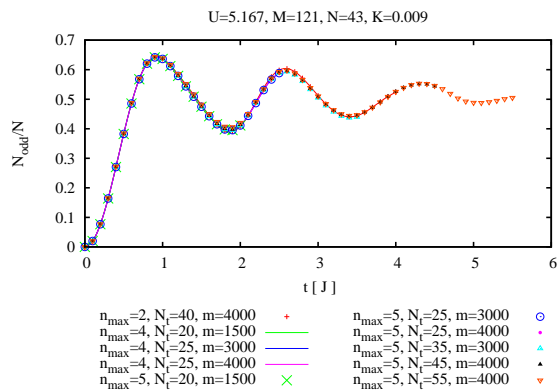}
\end{center}
\caption{Comparison of the time evolution of $\Nodd$. The first line (red) corresponds to a simulation, where at most two particles
are allowed per site. The other lines correspond to a maximal occupation of $\nmax=4$ and the symbols are the
corresponding data for $\nmax=5$. The number of of states per block is changed as described in the main text.
Note, the $N_t=55$ job did not finish due to a hardware failure, see main text.} \label{fig:NoddT}.
\end{figure}

In Fig.~\ref{fig:NoddT} we show results the occupation of the the odd sites, $\Nodd = \la \No \ra$,
$\No = \sum_{x=2y+1} \hat{n}_x$,
corresponding to the data corresponding to sample (c) in Fig.~(2)  of \cite{Trotzky:}.
We prepared our initial state by using a $J=0.4$ and a staggered potential $\hat{V}= V \sum_x (-1)^x \hat{n}_x$, $V=1$ in addition to an
on-site $U$ of 20.
Once we found the ground state we restarted the DMRG lowering $J$ to 0.1, and then $J=0$, performing five DMRG sweeps in each restart,
and performing $N_t=10$ time steps of $\Delta_t$. We then continued with $J=0$ for the initial state and increased the number of states
kept to 500, 750, 1500, 3000, 4000, performing 5 finite lattice sweeps for each restart, while at the same time we are increasing $N_t$ up to 25, 35, and 45.
We also report the result for $N_t=55$ time steps. However, this run was interrupted during the second sweep due to a hardware failure
which also destroyed the restart files. Therefore this data is not converged.

All these different runs display basically the same data, only the $\nmax=2$ run displays a slightly weaker damping of the oscillation of $\Nodd$.
In comparison to Fig.~(2c) of \cite{Trotzky:} 
we find a slightly smaller decay of the oscillations of $\Nodd$\footnote{Actually, in our simulation we find a different time scale
as compared to ref.~\cite{Trotzky:}. However, we didn't get a response from the authors whether we understood the parameter correctly.}
compared to the numerical simulations reported there, which can be attributed to the fact that there an ensemble average for different
particle numbers were reported, while here we only calculated the system with largest number of particles used in ref.~\cite{Trotzky:}.
More interestingly, in extending the simulation time beyond the one reported in  ref.~\cite{Trotzky:} we see a much smaller smaller decay
than the one reported in the experimental realization.

\begin{figure}
\begin{center}
\includegraphics[width=3in,clip=true]{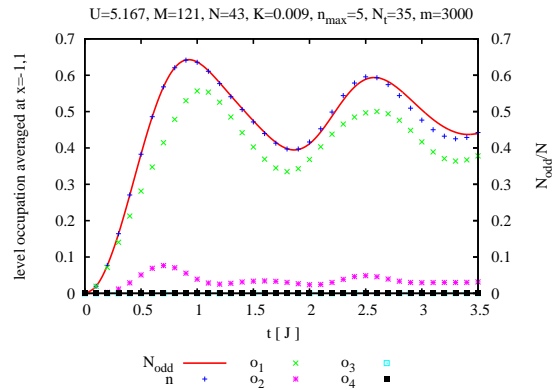}
\end{center}
\caption{Level occupancy of the odd sites in the centre of the system.
  The solid line corresponds to the average occupation of the odd sites, as shown in Fig.~\ref{fig:NoddT}.
  The 'plus' data show the average occupation of the two sites neighboring the center. 
  In addition we have resolved the particle occupation of these two sites is  resolved into the individual 
  level occupation of the single ($o_1$), double ($o_2$), triple ($o_3$), and quadruple ($o_4$) occupied levels;
  $n = \sum_{\ell=1}^5 \ell\,o_\ell$.} \label{fig:Bo}
\end{figure}

\begin{figure}
\begin{center}
\includegraphics[width=3in,clip=true]{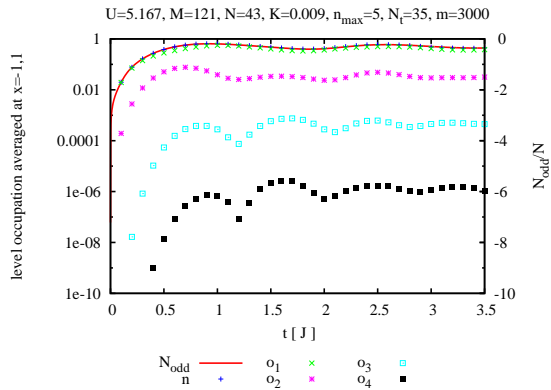}
\end{center}
\caption{Same data as in Fig.~\ref{fig:Bo} plotted on a logarithmic $y$-axis.} \label{fig:Bolog}.
\end{figure}

In Fig.~\ref{fig:Bo} provide the individual level occupation for the single ($o_1$), double ($o_2$), triple ($o_3$), and quadruple ($o_4$) occupied levels 
averaged over site $\pm 1$, where the complete expectation value of the local particle number is given by $n = \sum_{\ell=1}^5 \ell\, o_\ell$.
The result demonstrate that the particle number is already dominated by the single and double occupied level, the triple occupied level has
still some non-vanishing contribution, while the higher occupied levels do not contribute significantly to the particle number.
We also compare to the total average occupation $\Nodd / N$ which only shows a small deviation from the occupation of the two most inner odd sites.
Therefore, the boundary has only a small influence on the result.
We show the same data in Fig.~\ref{fig:Bolog} were we use a logarithmic scale on the $y$-axis in order to make the values of the triple and quadruple
occupied levels visible.

\begin{figure}
\begin{center}
\includegraphics[width=3in,clip=true]{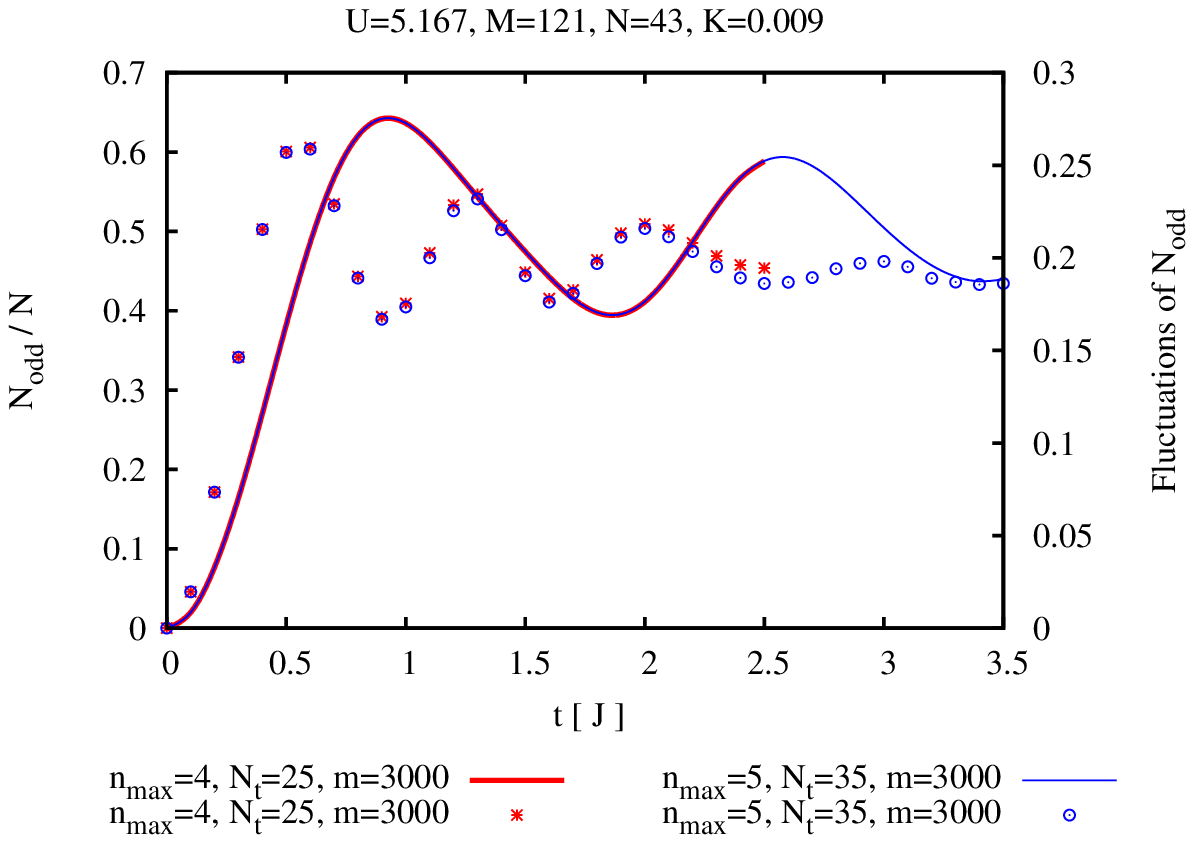}
\end{center}
\caption{The symbols show the average equal time fluctuations $ \bra{\Psi_t} \left(\No - \Nodd \right)^2\ket{\Psi_t} / N$,
       while the lines give the average occupation $\Nodd$ of the odd sites for comparison.} \label{fig:B_Flukt}.
\end{figure}

In contrast to the experiment the numerical simulations give us direct access to the measured quantities.
Specifically, in Fig.~\ref{fig:B_Flukt} we show the fluctuations $ \bra{\Psi_t} \left(\No - \Nodd \right)^2\ket{\Psi_t}$ of
the occupation. 
Actually, with the techniques described in ref.~\cite{Boulat-Saleur-Schmitteckert-2008,Branschaedel-Boulat-Saleur-Schmitteckert-2010,PS:PRL2011} the td-DMRG gives access to
frequency resolved noise correlations.
Within the time scale of our simulations they show basically the same decay as the oscillations for the average occupation,
supporting the idea of a relaxation process for the particles. However, from the numerics it is not clear, whether we see some kind of (local)
thermalization or just a dephasing of the dynamics. 
  
\begin{figure}
\begin{center}
\includegraphics[width=3in,clip=true]{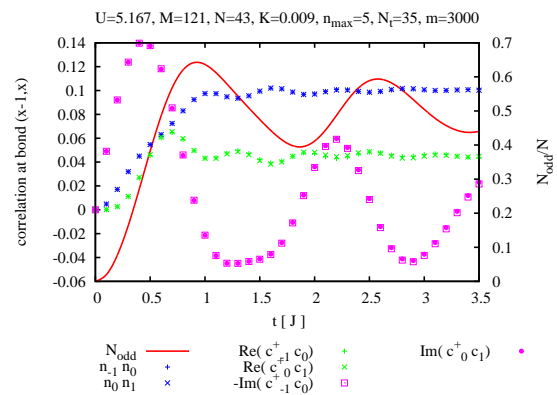}
\end{center}
\caption{Nearest neighbor correlations for the two inner bonds connected to the center site.
 The blue symbols denote the density density correlation $\la \hat{n}_{x-1} \hat{n}_x \ra$
   and the green symbol are the kinetic part ${\mathrm{Re}} \la \hat{a}^{+}_{x-1} \hat{a}^{}_x \ra$.
  Finally the circles correspond to bond currents ${\mathrm{Im}} \la \hat{a}^{+}_{x-1} \hat{a}^{}_x \ra$
  of the (0 -- 1) bond, and the squares to the negative of the (-1 -- 0) bond.
  The red line shows the average occupation of the odd sites $\Nodd/N$ for comparison.
 } \label{fig:Correllations}.
\end{figure}

In order to gain a deeper insight we look at the nearest neighbor correlation in the center of the trap.
In Fig.~\ref{fig:Correllations} we display the nearest neighbor density-density correlation  $ \bra{\Psi_t} \hat{n}_x \hat{n}_{x-1}c\ket{\Psi_t}$
and the nearest neighbor hopping element $ \bra{\Psi_t} \hat{a}^+_x \hat{a}^{}_{x-1} \ket{\Psi_t}$ for the two central bonds.
While the density-density correlation and the real part of the hopping element, i.e.\ the kinetic energy, show a similar
decay, the imaginary part of the hopping element, i.e.\ the bond currents, do not display such a decay during this short
time dynamics. Therefore, the nature of the relaxation, e.g.\ dephasing vs.\ thermalization is still an open question.

In summary we have shown that td-DMRG simulation can provide a deeper insight into the dynamics of strongly correlated Bose systems on a lattice.
In comparison to the experimental realization the numerics provides us with a large flexibility on the observables we want to look at.
Of course, by the mere definition of science the experiment describes nature correctly. However, if we want to simulate the dynamics of
the Hamiltonian eq.~\ref{eq:H} we claim that the numerics is currently not obsoleted by the experimental simulations.
Furthermore, the numerics allows to look at details, which are currently not all accessible by the experimental realization.
In addition we have pointed out the the nature of the relaxation process in these systems is still an open issue.
An interesting question that remains for future research is whether the (local) relaxation is due to a thermalization or
a dephasing process. This question bears similarity to the difference between the $\tau_1$ and $\tau_2$ relaxation in 
nuclear magnetic resonance measurements.
\acknowledgments
We would like to thank Avi Schiller and  Frithjof Anders for encouraging discussion at the FQMT'11.
Specifically the discussion of the connection of Martin John Rees's evening talk with Jens Eisert's presention
led to the research reported in this work. We would also like to thank V{\' a}clav {\v S}pi{\v c}ka for his tremendous work in organizing the FQMT'11 conference.

\end{document}